\documentclass[aps,prx,twocolumn,superscriptaddress,floatfix]{revtex4-1}

\usepackage{graphicx}% Include figure files
\usepackage{dcolumn}% Align table columns on decimal point
\usepackage{bm}% bold math
\usepackage{xcolor}
\usepackage[colorlinks=true, final]{hyperref}
\usepackage{titlesec}

\usepackage{wrapfig} % for figures in chapter introductions
\usepackage{relsize} % add to make math fonts smaller (corresponding command: \mathsmaller)
\usepackage{amsmath}

\usepackage{physics}

\usepackage[ulem=normalem,markup=underlined,final]{changes}
\definechangesauthor[color=red]{PVS}
\definechangesauthor[color=orange]{PVSn}
\definechangesauthor[color=blue]{RR}
\definechangesauthor[color=purple]{ML} 
\definechangesauthor[color=teal]{CH}
\newcommand{\note}[2][]{\added[#1,remark={#2}]{}}

\newcommand{\AlGaAs}{Al$_{0.33}$Ga$_{0.67}$As}
\newcommand{\mum}{$\mu$m~}

\newcommand{\be}{\hat b}
\newcommand{\bed}{\hat b^\dagger}
\renewcommand{\vec}[1]{\mathbf{#1}}

\usepackage{hyperref}

\begin{document}

%\title{Resonant-tunneling activated shallow impurities in GaAs quantum wells and their single-photon emission}% Force line breaks with \\
\title{Attractive dipolar coupling between stacked exciton fluids}% Force line breaks with \\

\author{Colin Hubert} 
\thanks{\normalsize{These authors have equal contributions}}

\affiliation{Paul-Drude-Institut f{\"u}r Festk{\"o}rperelektronik, Leibniz-Institut im Forschungsverbund Berlin e. V., Hausvogteiplatz 5-7, 10117 Berlin, Germany}
\email{hubert@pdi-berlin.de}

\author{Yifat Baruchi}
\thanks{\normalsize{These authors have equal contributions}}

\author{Yotam Mazuz-Harpaz}

\author{Kobi Cohen}
\affiliation{The Racah Institute of Physics, The Hebrew University of Jerusalem, Jerusalem 9190401, Israel}

\author{Klaus Biermann}
\affiliation{Paul-Drude-Institut f{\"u}r Festk{\"o}rperelektronik, Leibniz-Institut im Forschungsverbund Berlin e. V., Hausvogteiplatz 5-7, 10117 Berlin, Germany}

\author{ Mikhail Lemeshko}
\affiliation{Institute of Science and Technology Austria, Am Campus 1, 3400 Klosterneuburg, Austria}

\author{Ken West}
\author{Loren Pfeiffer}
\affiliation{Department of Electrical Engineering, Princeton University, Princeton, New Jersey 08544, USA}

\author{Ronen Rapaport}
\affiliation{The Racah Institute of Physics, The Hebrew University of Jerusalem, Jerusalem 9190401, Israel}

\author{Paulo Santos}
\affiliation{Paul-Drude-Institut f{\"u}r Festk{\"o}rperelektronik, Leibniz-Institut im Forschungsverbund Berlin e. V., Hausvogteiplatz 5-7, 10117 Berlin, Germany}

\date{\today}% It is always \today, today,

\begin{abstract}

The interaction between aligned dipoles is long-ranged and highly anisotropic\replaced[id=PVSn]{: it changes}{it changing} from repulsive to attractive depending on the relative positions of the dipoles. \replaced[id=PVSn]{We report on the observation of the attractive component of the dipolar coupling between  excitonic dipoles in stacked semiconductor bilayers.}{We report on the observation of attractive dipolar coupling between  excitonic dipoles in stacked semiconductor bilayers.} We show that the presence of a dipolar exciton fluid in one bilayer modifies the spatial distribution and increases the binding energy of excitonic dipoles in a vertically remote layer. The binding energy changes are explained by a many-body polaron model describing the deformation of the exciton cloud due to its interaction with a remote dipolar exciton. The results open the way for the observation of theoretically predicted new and exotic collective phases, the realization of interacting dipolar lattices in semiconductor systems as well as for engineering and sensing their collective excitations.

\end{abstract}

\maketitle

%%%%%%%%%%%%%%%%%%%%%%%%%%%%%%%%%%%%%%%%%%%%%%%%%%%%%%%%

\section{INTRODUCTION}

\replaced[id=PVSn]{The dipolar coupling normally dominates the }{The dipolar coupling is normally the dominant} interaction between charge-neutral species. The characteristic dipolar interaction energy between two dipoles with parallel axes and dipole moments ${\bf p_1}$ and ${\bf p_2}$ in a medium with dielectric constant $\epsilon$ can be expressed in the far field \deleted[id=PVSn]{approximation} as 

\begin{equation}
U_{dd}(\textbf{r}) = \frac{p_1p_2} {{4 \pi \varepsilon \varepsilon_0}} \frac{\left( 1-3 \cos^2\theta \right)}{r^3}
\label{Eq1}
\end{equation}

\noindent where \added[id=PVSn]{$\varepsilon_0$ is the vacuum permittivity,} $\theta$ is the angle  between  ${\bf p_1 || p_2}$ and \textbf r is the vector connecting the dipoles. While sharing the long decay range of the Coulomb interaction, the dipolar interaction is spatially anisotropic \replaced{and}{. In particular, it} changes from repulsive to attractive \deleted{when cos($\theta$)$ < \frac{1}{\sqrt{3}}$} at cos($\theta$)$ = \frac{1}{\sqrt{3}}$. In natural physical systems containing a large number of dipoles, this anisotropic character gives rise to complex phenomena including self organization, pattern formation, and instabilities in a wide range of dipolar fluids such as in ferro\deleted[id=CH]{- }magnetic or electric fluids~\cite{Lahaye_N448_672_07} as well collective effects in \deleted{ordered or disordered} dipolar lattices. \added[id=PVS]{Fascinating new phases of matter are expected if dipolar interactions are induced into quantum fluids, with an intricate interplay between the attractive and repulsive parts of the interaction and quantum mechanical effects. These new phases may have more than one continuous symmetry simultaneously broken, such as in the prediction of supersolidity.} Recent \replaced[id=RR]{experiments in superfluids of dilute cold  atomic species with magnetic dipoles have observed a non-isotropic gas expansion and an interaction-driven phase transition between a gas and a state of self-bound, self-ordered liquid droplets, stabilized by the balance between attraction and repulsion and quantum fluctuations}{dipolar effects have been identified to induce similar collective phenomena in superfluids of cold  atomic species with magnetic dipoles}~\cite{Lahaye_RPP72_126401_09,Ferrier-Barbut_PRL116_215301_16,Kadau_N530_194_16,Chomaz_PRX6_41039_16}.

\begin{figure}[tbhp]
\includegraphics[width = 1\columnwidth]{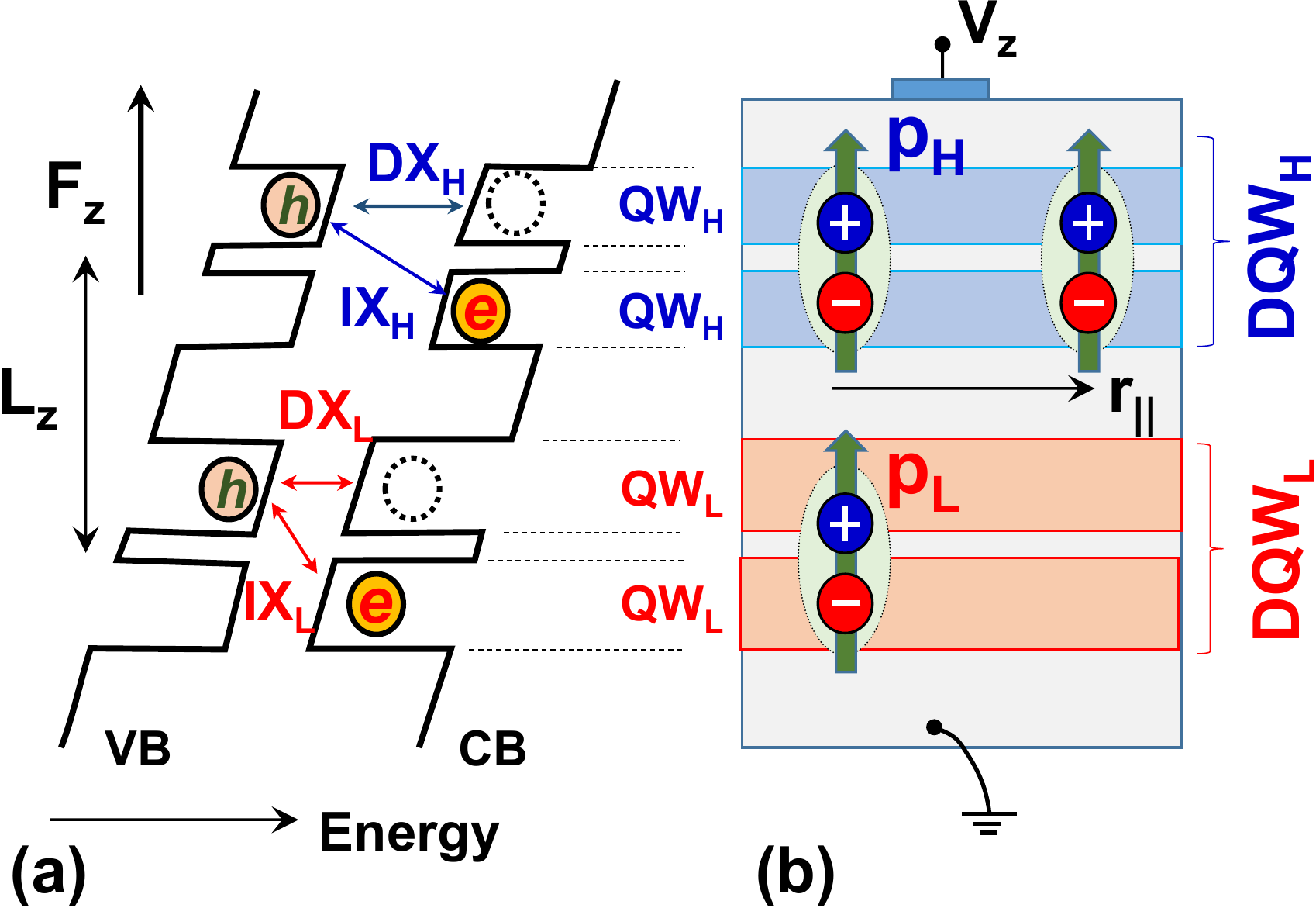}
\caption{Interactions between indirect (or dipolar) excitons (IX$_i$'s, i=L,H) in a sample with coupled double quantum wells (DQW$_i$'s).  (a) Energy diagram and (b) layer structure of the (Al,Ga)As sample. Each DQW consists of  two GaAs QWs separated by thin tunneling barriers. A 10~nm-thick (Al,Ga)As barrier between the DQWs prevents carrier tunneling between them. The transverse electric field, F$_z$, bends the conduction (CB) and valence bands (CB). Under laser excitation, the spatial separation between electrons (e) and holes (h) in each DQW creates IX$_i$s with an electric dipole moment, p$_i$, and reduced energy with respect to intra-well direct excitons (DX$_i$s). The DQWs are spaced by L$_z$ and have QWs with different widths to enable the selective optical excitation of their DX states  }
\label{fig:Potential}
\end{figure}
%\clearpage
%%%%%%%%%%%%%%%%%%%%%%%%%%%%%%%%%%%%%%%%%%%%%%%%%%%%%%%%%%%%%%%%%%%%%%%%%%%%%%%%%%%%%%%%%%%%%%
%(c) Intra (red curve) and inter-DQW (blue) dipolar interaction energy, U$_{dd}$(r), for a pair of IXs as a function of the lateral IX-IX separation r$_{||}$ calculated from Eq. (1) for the DQW stack used in the studies. While the intra-DQW interaction is always repulsive, the inter-DQW  one becomes attractive for $r_{||}<\sqrt{3}L_z$ and forms an IX-IX bound state with the indicated binding energy $|\Delta E_\mathrm{IX}|$. 

While cold magnetic atom experiments largely probe dipolar coupling in the regime of dilute quantum gases and small dipole moments,  fluids of electric dipoles in solid-state systems, and in particular\deleted{ cold,} spatially indirect  dipolar excitons (IXs)  in semiconductor bilayers, \deleted[id=CH]{have} open up opportunities to explore the \added[id=PVSn]{complementary} phase space of high density, large dipole-moments~\cite{Chen_PRB74_045309_06,Laikhtman_PRB80_195313_09,Shilo_NC4_2335_13,Misra_PRL120_47402_18,Stern_S343_55_14,Mazuz-Harpaz_arXiv1803_03818_18}.\note[id=RR]{Yotam Harpaz-Maxim Khodas-Ronen Rapaport Arxiv 2018 - donePVS} One interesting question is whether the attractive component of the dipolar interaction can be observed \added[id=PVSn]{and create self-bound states} in such solid-state systems. 
\replaced[id=PVSn]{ Access to this attractive component }{This observation} has been so far  impossible as all IX experiments have been conducted in a single dipolar bilayer of aligned dipoles, where the dipolar interaction is exclusively repulsive. 

In this work, we investigate the interaction between \replaced[id=PVSn]{mobile IX dipoles  confined in stacked bilayers. The bilayers are }{stacked bilayers of mobile dipoles. 
The studies were carried out with dipolar IXs confined in two stacked bilayers, each composed of } 
semiconductor double quantum wells (DQWs) (denoted as DQW$_L$ and DQW$_H$ in Fig.~\ref{fig:Potential}), each consisting of two quantum wells (QWs) separated by a thin barrier (i.e., with thickness smaller than the exciton Bohr radius, cf.~Fig.~\ref{fig:Potential}a,b). A vertical electric field applied across the structure  (F$_z$, cf. Fig.~\ref{fig:Potential}a) drives optically excited electrons and holes to different QWs while maintaining the Coulomb correlation between them. 
This  charge separation induced by F$_z$ imparts very long \replaced[id=PVSn]{lifetimes to the IXs, thus making them}{IX lifetimes reaching the $\mu$s range, thus making IXs} quasi-equilibrium excitations \deleted{, as well as a strong electric dipole moment to the IXs,} possessing a large dipole moment, which far exceeds the magnitude of atomic and molecular dipoles \deleted{and gives} thus giving rise to strong inter-particle interactions~\cite{Schindler_PRB78_045313_08, Laikhtman_PRB80_195313_09}. The intra-DQW repulsive component has received considerable experimental attention in IX systems and was utilized for many opto-electronic functional demonstrations~\cite{Schinner_PRL110_127403_13,Schinner_PRB83__11,Kouwenhoven_EEL18_607_92, Alloing_SR3_1578_13,High_S321_229_08,Kowalik-Seidl_NL12_326_12,High_OL32_2466_07,Grosso_NP3_577_09,Lundstrom_S286_2312_99,Krenner_NL8_1750_08,Bayer_PE12_900_02,Borges_PLA380_3111_16,Lacava__17_16}.  
Furthermore, \replaced[id=PVSn]{several}{many observations of} many-body collective effects related to the bosonic character of these interacting particles have been reported~\cite{High_Nature_12,Shilo_NC4_2335_13,Cohen_NL16_3726_16,Stern_PRL100_256402_08,Stern_S343_55_14,Alloing_E107_10012_14,Combescot_RoPiP80_66501_17,Dremin_POTSS46_170_04,Zhu_PRL74_1633_95,Misra_PRL120_47402_18}.\note[id=RR]{are all these on DQWs reporting condensation of sorts??? you can add Israel BJ PRL 2018 - done PVS}  

The stacked DQW structures  result in an attractive inter-DQW dipolar component  for small  lateral separation between the IXs, which has so far escaped experimental detection. 
Here, by using spatially-resolved spectroscopy, we show that the attractive component of the dipolar interaction induces density correlations  between IX fluids in remote DQWs, analogous to the remote dragging \cite{Narozhny_RMP88_25003_16} observed in solid-state electron-phonon, electron-electron \cite{Solomon_PRL63_2508_89}, and electron-hole \cite{Nandi_N488_481_12,Solomon_PRL63_2508_89} systems, but now involving  charge-neutral, bosonic species. Interestingly, the energetic changes induced by the remote dipolar coupling exceed the values predicted for formation of dipolar pairs \cite{PVS284}, and are non-monotonous in the fluid density. 
\replaced[id=PVSn]
{The large coupling energies, which  are attributed here to a self-bound, collective many-body fluid excitation identified as a dipolar polaron. The latter is analogous to self-bound three-dimensional entities with compensating attraction and repulsion like atomic nuclei, helium, and cold atom droplets.  The experimental findings demonstrate the feasibility of control and manipulation of dipolar species via remote dipolar forces. 
}
{The large coupling energies, which  are attributed here to a collective many-body fluid excitation identified as a dipolar polaron, demonstrate the feasibility of control and manipulation of dipolar species via remote dipolar forces. 
}
\added[id=PVS]{Furthermore,} the sensitivity to the fluid's local correlations opens new ways to study fundamental properties of correlated dipolar fluids.

\begin{figure}[tbhp]
\includegraphics[width = 1\columnwidth]{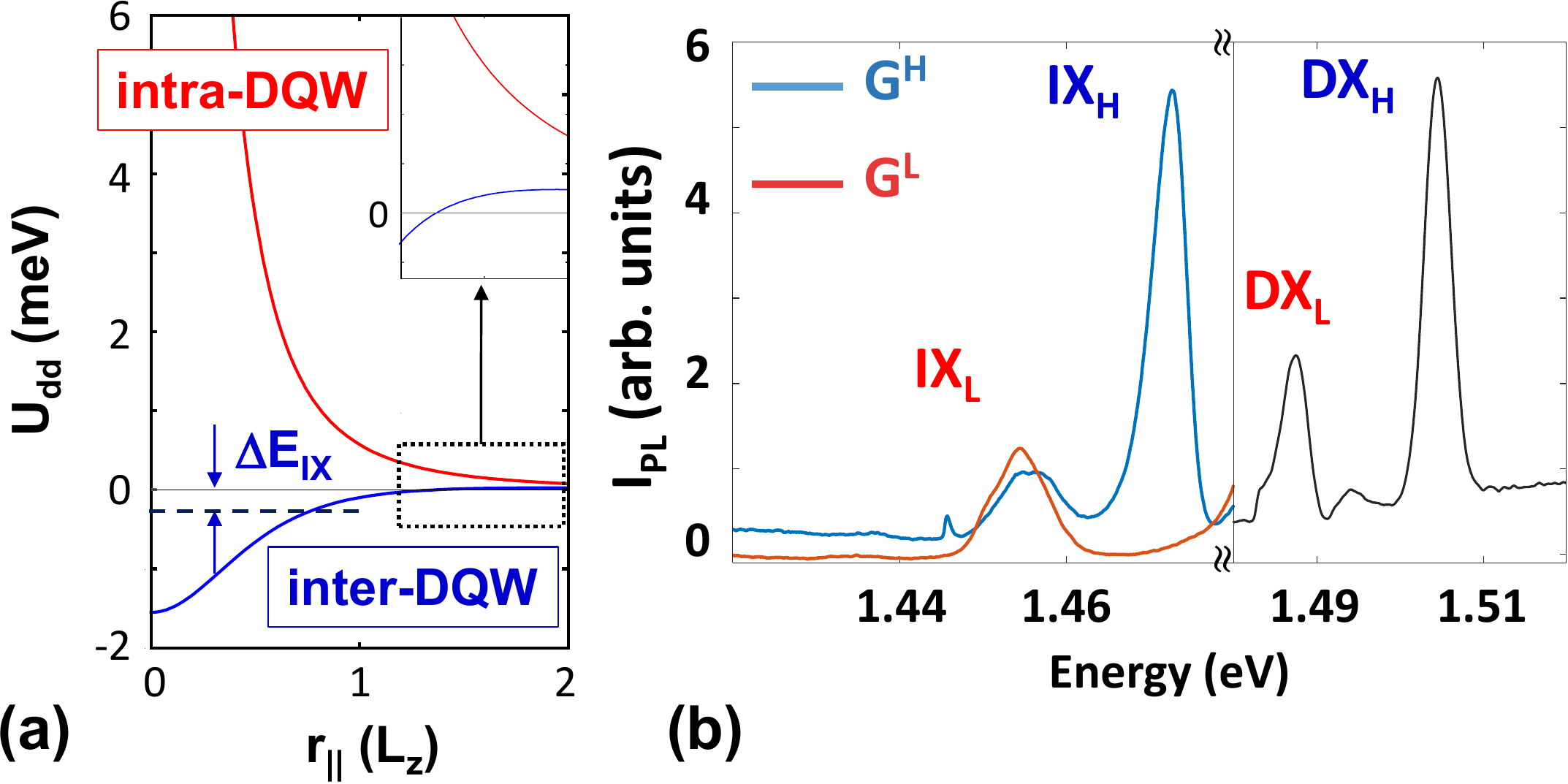}
\caption{Potential between indirect excitons and their excitation spectra (a) Intra (red curve) and inter-DQW (blue) dipolar interaction energy, U$_{dd}$(r), for a pair of IXs as a function of the lateral IX-IX separation r$_{||}$ calculated from Eq. (1) for the DQW stack used in the studies. While the intra-DQW interaction is always repulsive, the inter-DQW  one becomes attractive for $r_{||}<\sqrt{3}L_z$ and forms an IX-IX bound state with the indicated binding energy $|\Delta E_\mathrm{IX}|$. (b) Photoluminescence (PL) spectra of the  indirect (left panel) and direct (right panel) exciton transitions. Indirect excitons IX$_L$ and IX$_H$ (left panel) recorded under resonant excitation of their corresponding direct exciton transitions (right panel) using laser beams $G^L$ and $G^H$, respectively. 
}
\label{fig:Excitation}
\end{figure}
%\clearpage
%%%%%%%%%%%%%%%%%%%%%%%%%%%%%%%%%%%%%%%%%%%%%%%%%%%%%%%%%%%%%%%%%%%%%%%%%%%%%%%%%%%%%%%%%%%%%%

\section{EXPERIMENTAL CONCEPT}

The two closely spaced (Al,Ga)As DQWs are grown by molecular beam epitaxy (cf. Fig.~\ref{fig:Potential}(a,b)) on a GaAs (001) substrate. In order to enable selective optical excitation and detection, the DQWs (DQW$_L$ and DQW$_H$) have QWs of different thicknesses (QW$_L$ and QW$_H$), thus resulting in different resonance energies for their direct (DX$_i$) and indirect exciton (IX$_i$) transitions. Here, the subscripts $i = \mathrm{L,H}$ denote DQWs with the higher (H) and lower (L) excitonic energy. We will present experimental results recorded \added[id=PVS]{at 2~K} on two samples (samples A and B, details about both sample structures can be found in Appendix~\nameref{a:Samples}), both with QW widths of 10 and 12 nm for QW$_H$ and QW$_L$, respectively, and inter-QW spacing consisting of a 4~nm-thick \AlGaAs ~barrier. The 10~nm-thick \AlGaAs ~spacer layer between the DQWs prevents carrier tunneling, which would effectively result in the annihilation of the IXs.  Figure~\ref{fig:Excitation}(a) shows the intra- and inter-DQW dipolar potentials calculated for these structures using Eq.~\ref{Eq1}. \added[id=PVSn]{Note that the latter becomes attractive for small lateral separation between the particles.}

The two different QW thicknesses enable selective excitation and detection of IXs in each of the DQWs, as illustrated by the photoluminescence (PL) spectra of Fig.~\ref{fig:Excitation}(b) and \replaced[id=PVSn]{and the  excitation }{the} diagrams of Fig.~\ref{fig:SpatialDensityMaps}(a). A laser beam G$^{L}$ tuned to the DX$_L$ resonance only excites IX$_L$s in DQW$_L$ (throughout the paper superscripts $j=L,H,L+H$ denote excitation by laser beams G$^{L}$, G$^{H}$ and both, respectively). Since the DX$_L$ lies energetically below DX$_H$, a second laser G$^{H}$ tuned to DX$_H$ preferentially excites IX$_H$s in DQW$_H$ but also creates residual IX$_L$s in the neighboring DQW. One can, nevertheless, achieve a high excitation selectivity of IX$_H$s. In fact, from the  ratio between the PL intensities we estimated that G$^{H}$ excites DX$_H$ densities that are approximately 3.6 times higher than the DX$_L$ ones.

The PL experiments were carried out by exciting the sample with laser beams G$^{L}$ and G$^{H}$ with independently adjusted spot sizes and intensities (cf. Fig.~\ref{fig:SpatialDensityMaps}a). The interaction between the photo-excited exciton clouds was probed by mapping the PL intensities I$_{i}^{j}(x,y)$ with $\mu$m spatial resolution. The photo-excited IX densities, typically in the range between $10^{9}$ and $10^{11}~$cm$^{-2}$, were determined from the blue-shifts of the emission lines in the uncoupled systems after correction for correlation effects following the procedure depicted in Ref.~\cite{Laikhtman_PRB80_195313_09}  (cf. Appendix~\ref{a:IXDensity}). 

\section{EXPERIMENTAL RESULTS}

\subsection{Spatially resolved photoluminescence}

\begin{figure*}[t]
\centering
\includegraphics[width =1.9 \columnwidth]{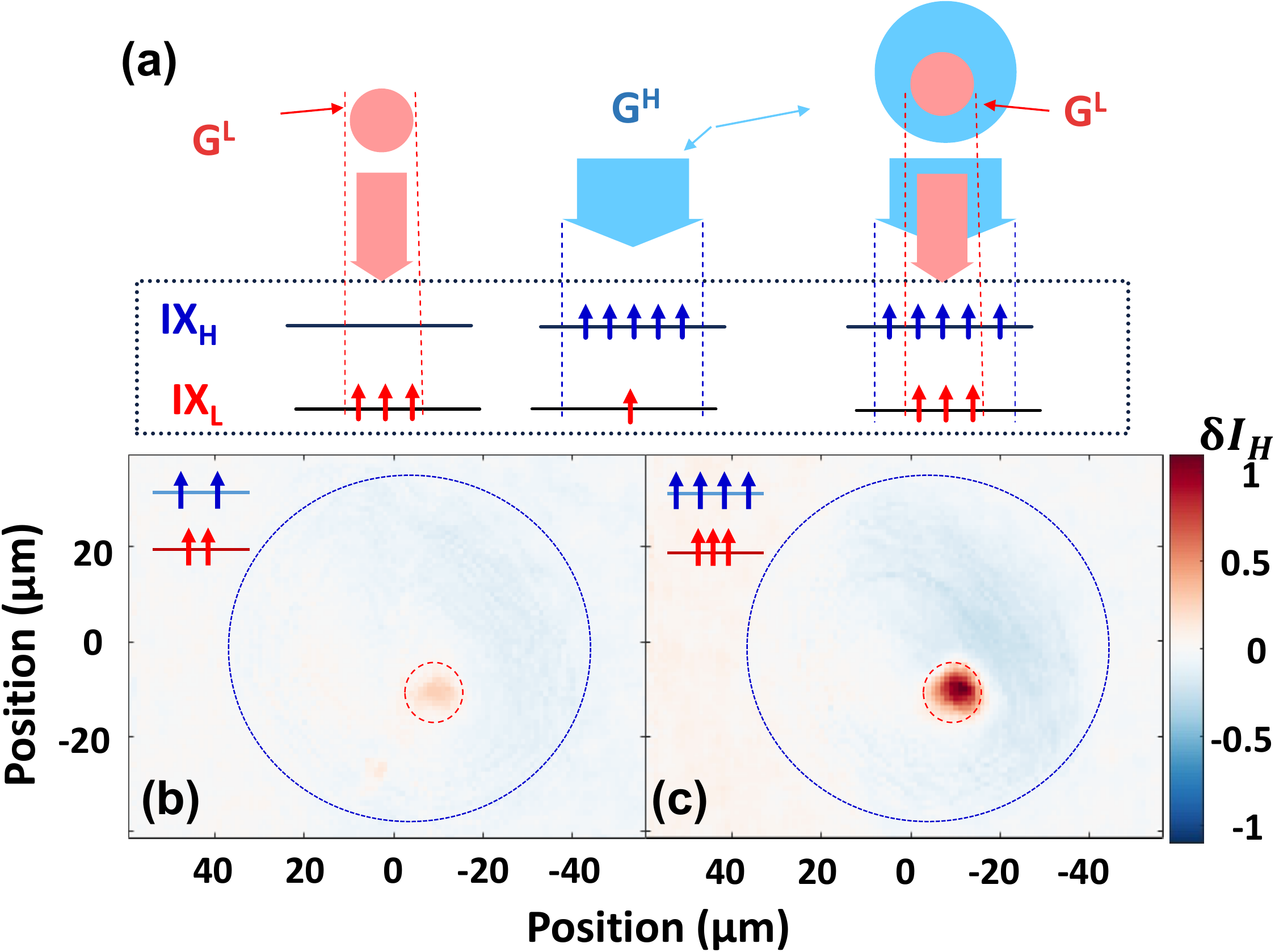}
\caption{Optical detection of inter-DQW interactions.
%The horizontal dashed lines mark the DX transitions. 
(a) Excitation schemes used in the experiments. IX clouds with different diameters are resonantly excited by lasers beams G$^{L}$ and G$^{H}$  tuned to the direct exciton transitions DX$_L$ and DX$_H$ of the DQWs. The emission from IX in the two DQWs is spectrally analyzed and detected with spatial resolution. (b,c) Maps of the relative change $\delta I_{H}(x,y)$ in the PL intensity of an IX$_H$ cloud induced by a narrow IX$_L$ cloud (marked by the dashed circle) for a fixed $G^L$ intensity and IX$_H$ densities at the center of $G^H$ of (b) $2.1\times 10^{10}$~cm$^{-2}$ and (c)  $4.6\times 10^{10}$~cm$^{-2}$, respectively.
}
\label{fig:SpatialDensityMaps}
\end{figure*}

The attractive inter-DQW interactions can be directly visualized by detecting intensity changes  $\Delta$I$_{i}$(x,y) in PL maps of a probing excitonic cloud in one of the DQWs induced by a perturbing cloud excited in the other DQW (cf.~Fig.~\ref{fig:SpatialDensityMaps}(a)). $\Delta I_{i}(x,y)$ is quantified according to:
\begin{equation}
\Delta I_{i}(x,y) = I_{i}^{L+H}(x,y) - \left[ I_{i}^{H}(x,y) + I_{i}^{L}(x,y)\right],\quad i=L, H
\label{Eq2}
\end{equation}

\noindent Here, the term within the brackets on the rhs accounts for the direct generation of IXs in the probing cloud by each of the laser beams. The most  sensitive approach to access inter-DQW interactions consists in detecting $\Delta I_{H}(x,y)$: since the perturbing laser $G_L$ does not directly excite IX$_H$, one obtains $\Delta I_{H}(x,y) \approx I_{H}^{L+H}(x,y) - I_{H}^{H}(x,y)$.
%\noindent Here,  the last term on the rhs corrects for PL from DQW$_i$ arising from the resonant laser excitation of the other DQW$_{\bar i}$. This term essentially vanishes for an IX$_H$ cloud, which cannot be excited by a laser tuned to the lower energy IX$_L$ state,  thus making  $\Delta I_{H}(x,y)$ a very sensitive probe of inter-DQW interactions.
%

%%%%%%%%%%%%%%%%%%%%%%%%%%%%%%%%%%%%%%%%%%%%%%%%%%%%%%%%%%%%%%%%%%%%%%
% Figure 2
%%%%%%%%%%%%%%%%%%%%%%%%%%%%%%%%%%%%%%%%%%%%%%%%%%%%%%%%%%%%%%%%%%%%%%

Figure~\ref{fig:SpatialDensityMaps}(b,c) displays a map of the relative changes  $\delta I_{H}(x,y)=\Delta I_{H}(x,y)/I_{H}^{H}(x,y)$ in PL intensity of an extended IX$_H$ probing cloud induced by a perturbing IX$_L$ cloud \replaced[id=PVSn]{in sample A.}{. The experiments were carried out on Sample A.} The probing cloud has %an approximately Gaussian shape with 
a diameter of 60~$\mu$m (cf. blue dashed circle), 
\replaced[id=PVSn]
{while the perturbing  $G^L$ beam excites a $20~\mu$m-wide IX$_L$ cloud with a density of approximately $1.1\times 10^{10}$~cm$^{-2}$ at its center  (cf. red dashed circle).}
{while the perturbing cloud has a smaller diameter of $20~\mu$m (cf. red dashed circle).}
%The perturbing \replaced[id=CH]{beam $G^L$ excites a cloud of IX$_L$ with a density of}{of IX$_L$ has a constant density %of} approximately $1.1\times 10^{10}$~cm$^{-2}$ \deleted[id=PVS]{, taken} at its center.
%%\added[id=CH]{, as measured for the isolated case}.  
This perturbing IX$_L$ cloud induces a local increase in the IX$_H$ density. The IX optical cross-section is negligibly small, so that IXs are created by first creating a DX, then converting to an IX. Thus the perturbing laser $G^L$  beam effectively does not excite IX$_H$s (cf. Fig.~\ref{fig:Excitation}b), and the enhanced emission provides a direct evidence for an attractive IX$_H$-IX$_L$ inter-DQW coupling. Furthermore, as the IX lifetime within the probing cloud is not expected to \replaced[id=PVS]{change appreciably under the}{depend on the} perturbing beam, \replaced[id=PVS]{one can assume  the relative density changes $\delta n_H(x,y)$  to be approximately equal to  $\delta I_\mathrm{PL,H}\left(x,y\right)$.}{one can safely assume  the relative density changes $\delta n_\mathrm{H}(x,y)$  to be equal to $\delta I_\mathrm{PL,H}\left(x,y\right)$.} 

%xxThe maps were recorded for increasing IX$_H$ densities at the center of the cloud (from $2.1\times 10^{10}~cm^{-1}$ in \ref{fig:SpatialDensityMaps}(c) to $4.6\times10^{10}~cm^{-1}$ in \ref{fig:SpatialDensityMaps}(e)) while holding the density of IX$_L$s constant at $\approx 1.1\times 10^{10}~cm^{-2}$.
% 

The emission from the probing cloud at the overlapping region of the beams 
enhances significantly with the IX density. Figure~\ref{fig:SpatialDensityMaps}d displays a PL map recorded by increasing the intensity of $G^H$\deleted[id=CH]{ to yield an  IX$_H$  density of $\approx 4.6\times 10^{10}~cm^{-2}$} (note that the density of the perturbing cloud also increases due to the absorption  of $G^H$ photons in DQW$_L$ \added[id=PVS]{, cf. Fig.~\ref{fig:SpatialDensityMaps}}). Under the higher IX densities, \replaced[id=PVS]{the PL intensity from the IX$_H$ cloud doubles}{the relative PL change $\delta I_{H}(x,y)$ reaches up to 210\%} \added[id=CH]{in the region of the perturbing beam}.

%%%%%%%%%%%%%%%%%%%%%%%%%%%%%%%%%%%%%%%%%%%%%%%%%%%%%%%%%%%%%%%%%%%%%%
% Figure 3
\begin{figure}[tbhp]
\centering
\includegraphics[width = 1\columnwidth]{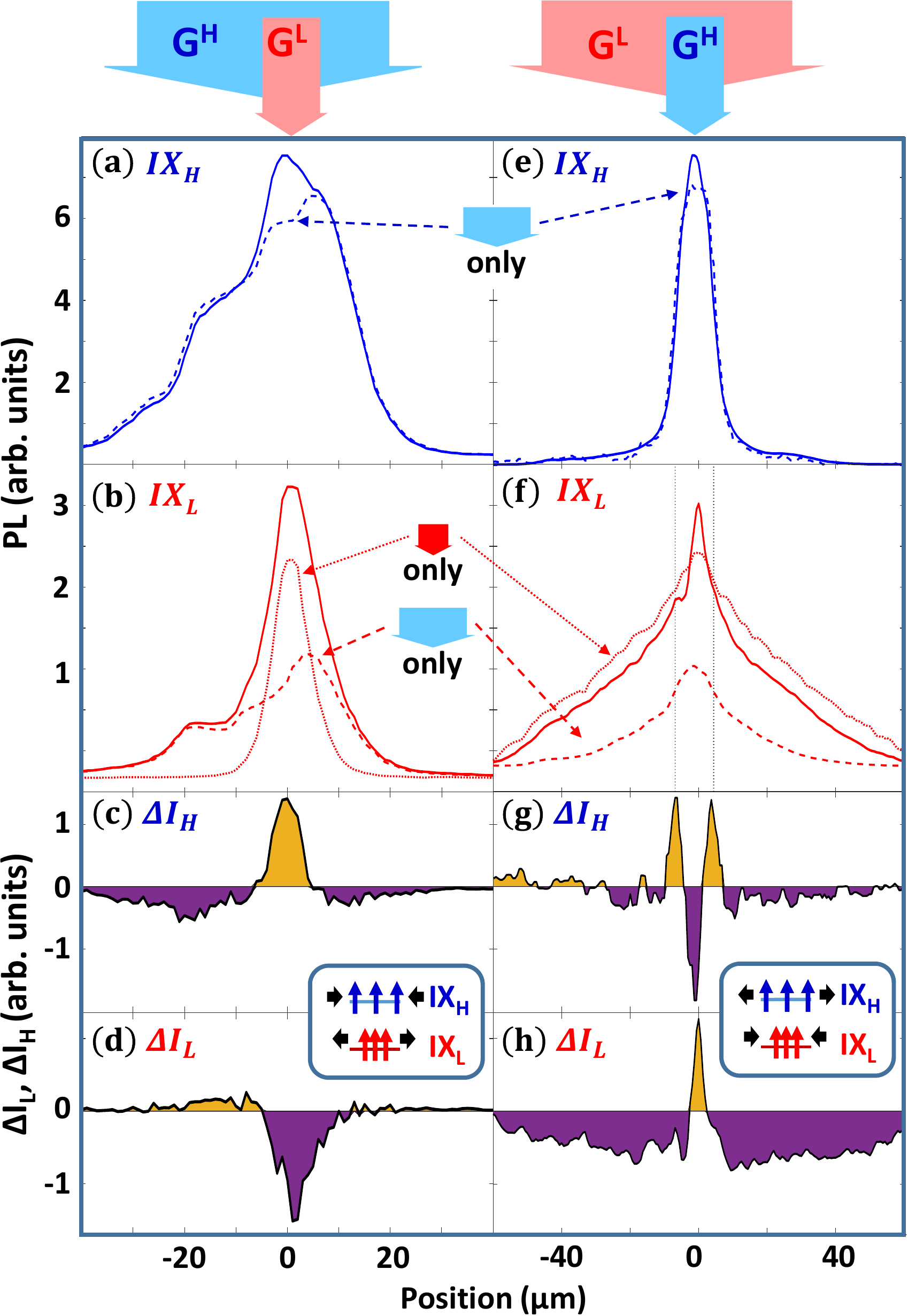}

\caption{Spatially resolved \replaced[id=PVSn]{PL}{photoluminescence} intensity profiles recorded using the configurations of laser beams sketched in the top for the (a)--(d)  IX$_H$ reservoir and (e)--(h)  IX$_L$ reservoir. In the notation I$_{i}^{j}(x,y)$,  the subscript denotes the probed species ($i=L, H$), while the superscript defines the laser excitation conditions (i.e., $j=L,H,L+H$ for excitation by laser beams G$^{L}$, G$^{H}$ and both, respectively). 
(c)-(f) Interaction-induced changes of the IX$_L$ ($\Delta I_L$) and IX$_H$ ($\Delta I_H$) photoluminescence determined according to Eq.~\ref{Eq2}. The inset depicts the expected forces \replaced[id=PVS]{exerted on}{on the populations, given the attractive force between} the two clouds. The left (right) panels were recorded on sample A (B) using beam widths of 20~\mum (6.5~\mum) and 60~\mum  (5.5~\mum) for G$^{L}$ and  G$^{H}$, respectively. \deleted[id=PVSn]{All profiles were recorded across the interaction region defined by the intersection of the two laser spots.}
}
\label{fig:SpatialCrossSections}
\end{figure}
%%%%%%%%%%%%%%%%%%%%%%%%%%%%%%%%%%%%%%%%%%%%%%%%%%%%%%%%%%%%%%%%%%%%%%

Further insight into the inter-DQW interaction can be gained from cross-sections  of the PL images across the overlap region of the two 
\replaced[id=PVS]{clouds, as illustrated in Fig.~\ref{fig:SpatialCrossSections}.  The left panels correspond to the experimental configuration of Figs.~\ref{fig:SpatialDensityMaps}c-d with a wide IX$_H$ and a narrow  IX$_L$ cloud (cf. diagrams in the upper part of the figure). The changes in the  IX$_H$ emission in Fig.~\ref{fig:SpatialCrossSections}(a) reproduce the density enhancement within the overlap area of the laser beams.}{clouds, as illustrated in the left panels of Fig.~\ref{fig:SpatialCrossSections}. The changes in the  IX$_H$ emission in Fig.~\ref{fig:SpatialCrossSections}(a) reproduce the density enhancement within the overlap area of the laser beams (cf. diagrams in the upper part of the figure).} 
The corresponding differential profile  ${\Delta I}_{H}$ in Fig.~\ref{fig:SpatialCrossSections}(c) shows that the enhanced concentration of IX$_H$ within the overlap region is accompanied  by a depletion around it. This behavior follows from the fact that the perturbing $G^L$ beam does not change the overall IX$_H$ density. As a consequence, the enhanced concentration at the overlap area must then arise from the IX$_H$ flow from the surrounding areas.
\note[id=RR]{the colors in Fig. 3c,d,g,h are confusing. we should have one color for a positive change and one color for a negative change in all of them consistently. They should not be red and blue to not confuse with the different exciton color codes.}

The attractive force leading to the enhanced IX$_H$ density should be accompanied by a back-action force on the perturbing IX$_L$ cloud (cf. inset of Fig.~\ref{fig:SpatialCrossSections}(c))). In order to extract information about this back-action effect on the IX$_L$ profiles, one needs to  account for the fact that IX$_L$s are also excited by the $G^H$ beam (cf.  Figs.~\ref{fig:SpatialDensityMaps}(a) and \ref{fig:SpatialCrossSections}(b)), thus leading to  a non-vanishing $I_{L}^{H}(x,y)$ term on the rhs of Eq.~\ref{Eq2}. The intensity variation ${\Delta I}_{L}(x,y)$ calculated from this equation and displayed in Fig.~\ref{fig:SpatialCrossSections}d shows indeed a depletion of the IX$_L$ density around the beam overlap region induced by the remote interaction.

The reciprocal of the above effect is expected if the previous experiment is carried out using a narrow $G^H$ spot to perturb an IX$_L$ cloud excited by an extended $G^L$ beam. Qualitatively similar results were indeed obtained in this situation, as illustrated by the right panels of Fig.~\ref{fig:SpatialCrossSections} (here, \replaced[id=PVSn]{smaller laser spots relative to the right panel}{relatively small laser spots} were employed with diameters of  6.5~\mum and 5.5~\mum for $G^L$ and $G^H$, respectively). %were selected to enhance the impact of inter-DQW interactions on the IX drift and diffusion. One example is illustrated in Fig.~\ref{fig:SpatialCrossSections}(e): here, the attractive interaction between the two clouds does not enhance the IX$_L$ density at the center of their overlap area but suppresses the IX$_L$ out-diffusion away from it.
Since the mobility of IX$_L$ is much larger than that of the IX$_H$\cite{Voros_PRL94_226401_05}, the density  disturbance of the IX$_L$  is far more extended than that of the IX$_H$, as is seen from the comparison of Fig.~\ref{fig:SpatialCrossSections}(c,g) to Fig.~\ref{fig:SpatialCrossSections}(d,h).  

%%%%%%%%%%%%%%%%%%%%%%%%%%%%%%%%%%%%%%%%%%%%%%%%%%%%%%%%%%%%%%%%%%%%%%
% Figure 5
\begin{figure}[htbp]
\includegraphics[width =1\columnwidth]{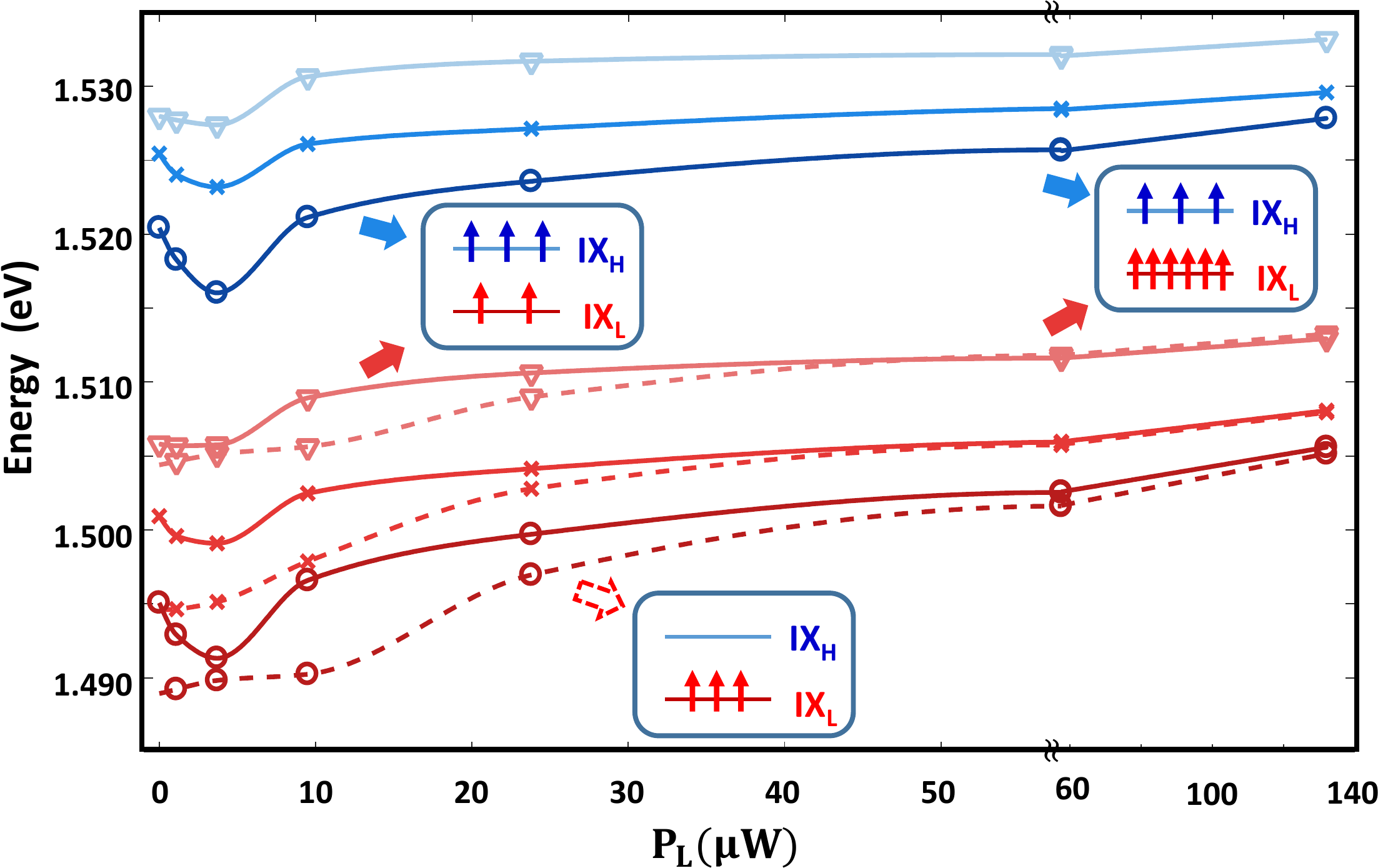}
\caption{IX resonance energies  as a function of the G$^L$ excitation power recorded for a constant G$^H$  power density of $3.6~$W/cm$^{2}$. The symbols correspond to electric fields of 19~kV/cm (triangles), 25.3~kV/cm (crosses), and 28.5~kV/cm (circles) applied in reverse bias across sample B. The solid lines were recorded in the presence of both the IX$_H$ and IX$_L$  clouds,  while the dashed lines  show the IX$_L$s's energies detected in the absence of the IX$_H$ cloud. The lines connecting the measurements (symbols) are spline interpolations.}
\label{fig:SpectralResults}
\end{figure}
%%%%%%%%%%%%%%%%%%%%%%%%%%%%%%%%%%%%%%%%%%%%%%%%%%%%%%%%%%%%%%%%%%%%%%
\subsection{Exciton binding energy}

The attraction between the remote IX clouds should be accompanied by changes in the observed IX energies within the overlapping regions of the two beams. The solid lines in Fig.~\ref{fig:SpectralResults} summarize the dependence of the IX$_L$ (lower curves) and IX$_H$ (upper curves) energies recorded in sample B by fixing the IX$_H$ density and progressively increasing the density of IX$_L$ species (stated in terms of the $G^L$ laser flux). The different  curves correspond to different electric fields applied across the structure. The latter controls the IX energies as well as the IX densities in both DQWs (larger electric fields correspond to larger steady state densities for the same excitation power \cite{Mazuz-Harpaz_PRB95_155302_17}).

%We first concentrate on the energies for IX$_L$ (lower curves). Here, the dashed lines display reference profiles for the dependence of the IX$_L$ energy on excitation power measured in the absence of IX$_H$'s.  The later show the characteristic blue shift associated with the repulsive intra-DQW IX-IX interactions. When the B$^{(H)}$ is turned on (solid lines), the IX$_L$ energy blue-shifts relative to the previous case and also develops a pronounced minima for B$^{(L)}$ powers between 0 and 10~$\mu$W. Similar minima are also observed in the energy dependence of the IX$_H$ resonances.

For all applied fields, the energies of both the IX$_L$ and IX$_H$ resonances show a pronounced minimum for G$^L$ powers between 0 and 10~$\mu$W followed by a smooth increase in energy for higher IX$_L$ excitation powers.  Note that for a given applied field, the IX$_H$ density remains constant as the IX$_L$ density changes. Strikingly, the minima only appear when both species are present \added[id=PVSn]{and have similar amplitudes for  IX$_L$ and IX$_H$}. In fact, the energy profiles for the IX$_L$ species recorded under resonant excitation by solely $G^L$ (dashed lines) show only the characteristic energy increase associated with the repulsive intra-DQW IX-IX interactions. 
The reduction in the excitonic resonance energies is attributed to the attractive inter-DQW interactions, which display a \deleted[id=CH]{non-trivial} non-monotonic density dependence. They appear for G$^L$ laser powers within a relatively small range and essentially vanishes at high IX$_L$ densities, where the IX energy becomes equal to the uncoupled case (dashed lines).

\subsection{The dipolar-polaron model}

\begin{figure}[htbp]
\includegraphics[width = 1\columnwidth]{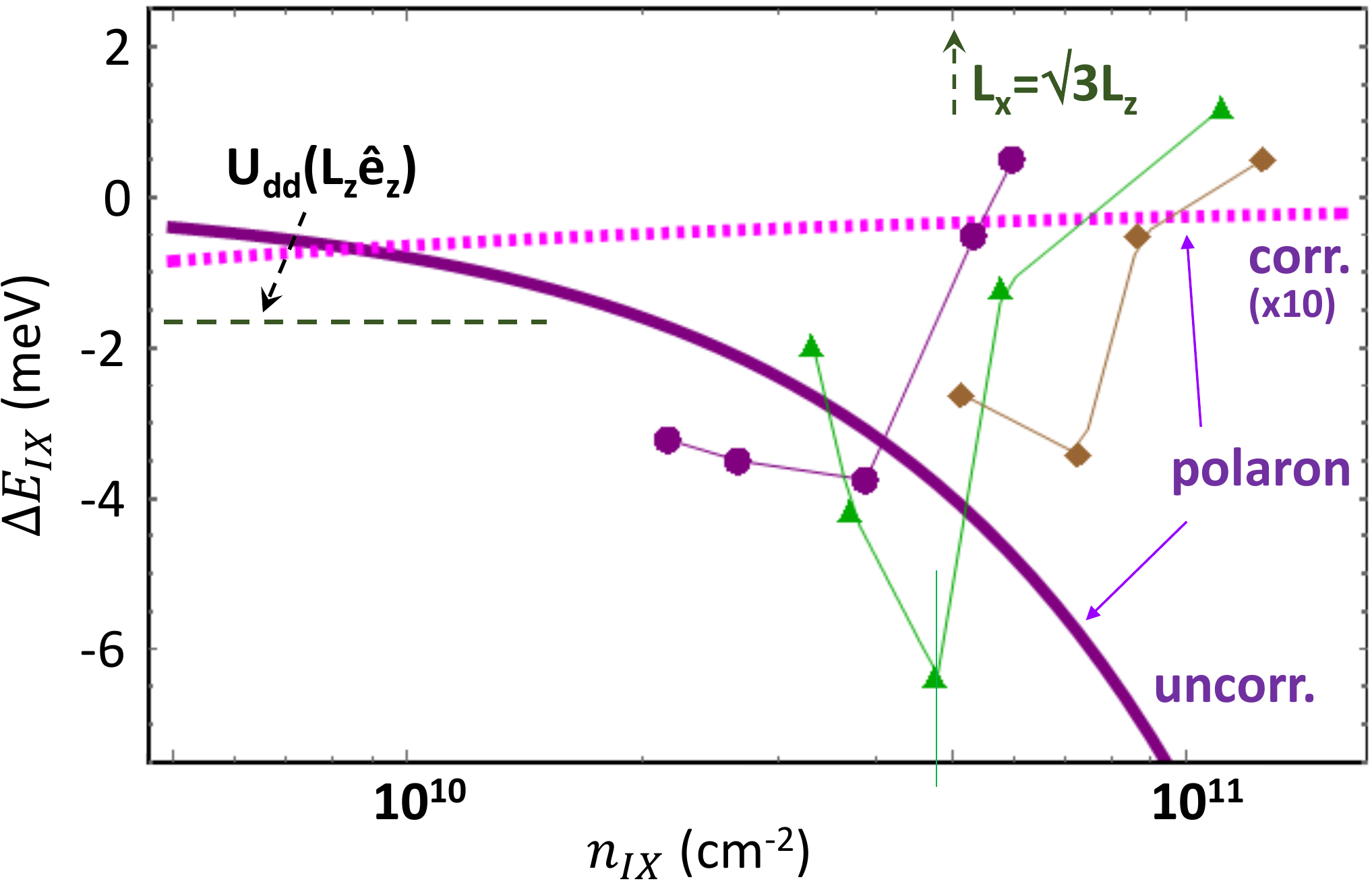}
\caption{Interaction-induced  energy shifts $\Delta E_\mathrm{IX}$ of IX$_H$ excitons induced by a remote IX$_L$ cloud with different densities $n_\mathrm{IX}$.
Results are shown for probing IX$_H$ densities of $8\times 10^{10}$~cm$^{-2}$ (dots), $9\times 10^{10}$~cm$^{-2}$ (triangles) and $9.7\times 10^{10}$~cm$^{-2}$ (diamonds). The error bar only shown for $n_\mathrm{IX}=5\times 10^{10}$~cm$^{-2}$ applies for all data points. The vertical dashed arrow marks the density for which $L_x=\sqrt{3}L_z$. 
The thick dotted and solid lines display the prediction of the polaron model in the limit of fully correlated (corr., cf.~Eq.~\ref{EqE02}) and uncorrelated (uncorr., cf.~Eq.~\ref{E03}) IX gases, respectively. This model is sketched in the inset, where the thick arrows schematically represent the distortion of the IX cloud. 
The horizontal dashed line marks the minimum of the inter-DQW interaction potential $U_{dd}(L_z{\hat{e}_z}+r_{||}\hat{e}_{||})$ given by Eq.~(\ref{Eq1}).}
\label{fig:Theory}
\end{figure}

The experiments described above provide evidence for an attractive dipolar interaction between IX clouds located in stacked DQWs. 
\replaced[id=PVSn]
{
The inter-DQW interaction also induces density-dependent energetic shifts (cf.~Fig.~\ref{fig:SpectralResults}), which will be quantified by an inter-DQW binding energy  $\Delta E_\mathrm{IX}$ defined as the difference between the IX energies with and without inter-DQW interactions, both  referenced at the same IX density. }
{The inter-DQW interaction also induces density-dependent shifts $\Delta E_\mathrm{IX}$ of the  IX energy (cf.~Fig.~\ref{fig:SpectralResults}). }
\replaced[id=PVSn]
{The dependence of  $\Delta E_\mathrm{IX}$  for the IX$_H$ cloud on the perturbing IX$_L$ density $n_\mathrm{IX}$ are summarized in Fig.~\ref{fig:Theory}. 
The $n_\mathrm{IX}$ values for the different $G^L$ laser powers and applied fields were extracted from the data in Fig.~\ref{fig:SpectralResults} following the procedure delineated in Appendix~\nameref{a:IXDensity}. 
}
{The IX$_L$ densities $n_\mathrm{IX}$  for the different $G^L$ laser powers and applied fields were extracted from the data in Fig.~\ref{fig:SpectralResults} following the procedure delineated in Materials and Methods. 
The dependence of  $\Delta E_\mathrm{IX}$  for the IX$_H$ cloud on the perturbing IX$_L$ density $n_\mathrm{IX}$ are summarized in Fig.~\ref{fig:Theory}.
}

%%%%%%%%%%%%%%%%%%%%%%%%%%%%%%%%%%%%%%%%%%%%%%%%%%%%%%%%%%%%%%%%%%%%%%
% Figure 7
%%%%%%%%%%%%%%%%%%%%%%%%%%%%%%%%%%%%%%%%%%%%%%%%%%%%%%%%%%%%%%%%%%%%%%

The three sets of experimental data points in Fig.~\ref{fig:Theory} correspond to the three different fixed IX$_H$ densities extracted from the data sets of Fig.~\ref{fig:SpectralResults} for the three different applied electric fields (the \added[id=PVSn]{associated} probing IX$_H$ densities are  \replaced[id=PVSn]{listed in the figure caption}{$8.0\times 10^{10}$~cm$^{-2}$, $9.0\times 10^{10}$~cm$^{-2}$ and $9.7\times 10^{10}$~cm$^{-2}$}). 

Surprisingly, the maximal observed energy shifts are very large \replaced[id=PVSn]{reaching up to}{, ranging between 4 to} 7~meV. Such large energies are not expected if one considers only the mutual attractive interaction and binding of a pair of IXs, one from each DQW layer. The formation of such bound pairs (``vertical IX molecules'') was recently investigated theoretically by Cohen and co-workers~\cite{PVS284}. The  inter-DQW dipolar potential calculated \replaced[id=PVSn]{ for }{by applying their approach to} the structures investigated here is illustrated in Fig.~\ref{fig:Potential}(c). This attractive potential binds the two IX species into an IX ``molecule'' with a binding energy $\Delta E_\mathrm{IX}$ of only a few tenths of a meV (dashed line in  Fig.~\ref{fig:Potential}). $\Delta E_\mathrm{IX}$ is much smaller than the depth of the potential due to the large zero-point energy corrections arising from the small (reduced) mass of the particles and short spatial extent of the potential. The measured IX energy shifts in Fig.~\ref{fig:Theory} are over an order of magnitude larger than the estimated IX molecular binding energy.  These shifts are also significantly larger than the depth of the attractive inter-DQW potential of Fig.~\ref{fig:Potential}(c), which is indicated by the horizontal dashed line in Fig.~\ref{fig:Theory} (see a more detailed analysis in Appendix~\nameref{a:Electrostatic}). 

This disagreement between the calculated molecular IX binding energies and the experimental values is not unexpected, since 
the large energetic shifts appear for rather high IX densities, for which the  average lateral inter-particle separation within each layer ($L_x$) becomes  comparable to the vertical separation ($L_z$) between the DQWs. Under these conditions many-body interactions can no longer be neglected. We therefore consider the mutual deformation of the exciton clouds induced by inter-DQW interactions, which may lead to the  formation of an IX dipolar-polaron.  For simplicity, we consider the case where the density in one of the layers is low, so that we can approach the problem as an ``impurity problem'': a single IX in  DQW$_2$ interacting with an exciton fluid in DQW$_1$ (cf. inset of Fig.~\ref{fig:Theory}). This \replaced[id=PVS]{approximation}{approach}, which is described in detail in the Sec.~SM4, might still qualitatively capture the case of large IX densities in both layers. We start from a Fr\"ohlich-type polaron Hamiltonian~\cite{Devreese15}:

\begin{equation}
\label{H1}
\hat H = {\frac{\hat{\vec{p}}^2}{2 M}}  + \sum_{\vec k} \added[id=ML]{\hbar \omega (k)} \bed_\vec{k}  \be_\vec{k} +  \sum_\vec{k} U (k) (e^{-i \vec{k} \hat {\vec{r}}} \bed_\vec{k}  +e^{i \vec{k} \hat {\vec{r}}} \be_\vec{k} ),
\end{equation}

\noindent where $\sum_{\vec k} = (2 \pi)^{-2} \int d^2k$. The first term describes the ``impurity'' (i.e., the single IX in DQW$_2$) with momentum $\hat{\vec{p}}$ and mass $M$ while the second term gives the kinetic energy of the bosonic bath (e.g.\ phonons in the exciton liquid formed in DQW$_1$), parametrized by the dispersion relation $\omega (k)$. The last term gives the impurity-boson interactions. Here, $U (k) = f(k) V(k)$, where $V(k)$ is the Fourier transform of the two-body interaction potential $U_{dd}$ in Eq.~\ref{Eq1}  and $f(k)$ is a function that depends on the correlation state of the IX gas (cf. Eq.~\ref{EqVk} of SM\deleted[id=PVSn]{, Sec.~\ref{SM_polaron_model}}).
 
\begin{figure}[htbp]
\includegraphics[width = 1\columnwidth]{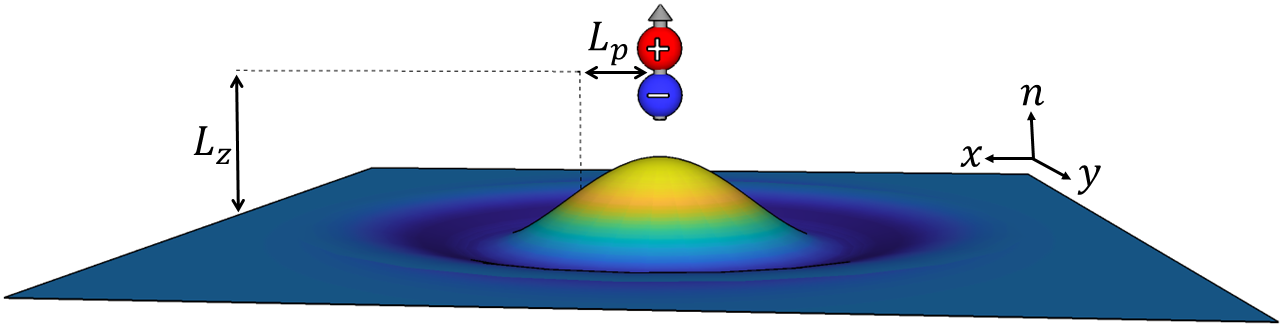}
\caption{The polaron model presented here assumes a single IX in the upper layer, interacting with a bath of IXs in the lower layer. The presence of this exciton causes changes in the density distribution of the IX fluid, which can be described as coupled acousto-electric waves, or polarons. The breadth of the polaron, L$_p$, is determined by the strength of the inter-layer dipolar coupling, which itself is highly dependent on the separation between layers, L$_z$.}
\label{fig:Concept}
\end{figure}

If we consider a static impurity (an ``infinite-mass polaron'', $M=\infty$, located at $\vec{r} = 0$), the Hamiltonian in Eq.~\ref{H1} can be diagonalized using a coherent-state transformation (see details in Sec.~SM4), yielding  
a negative  ``deformation energy'' $\Delta E_\mathrm{IX}$, as depicted in Fig.~\ref{fig:Concept}. 
In order to quantitatively estimate $\Delta E_\mathrm{IX}$,  we analyze two limiting solutions of Eq.~\ref{H1} depending on the correlation state of the IX fluid. We  first consider a gas of non-interacting IXs with  dispersion relation given by 
$\hbar \omega(k) \equiv \varepsilon(k) = \hbar^2 k^2/(2m)$, where $m = m_e + m_{hh}$ is the exciton mass (we take $m_e = 0.067$ and $m_{hh} = 0.23$ for the electron and in-plane heavy-hole effective masses in GaAs). In this case,  the energy shift becomes:

\begin{equation}
\label{EqE02}
%\Delta E_\mathrm{IX}  =  - n_\mathrm{IX} \mu_1^2 \mu_2^2 \frac{\pi m}{L_z^2}
\Delta E_\mathrm{IX}  =  - n_\mathrm{IX} \mu_1^2 \mu_2^2 \frac{\pi m}{\added[id=ML]{\hbar^2} L_z^2},
\end{equation} 

\noindent where $\mu_i=p_i/\added[id=ML]{\sqrt{4 \pi \varepsilon \varepsilon_0}}$. 

%%%%%%%%%%%%%%%%%%%%%%%%%%%%%%%%%%%%%%%%%%%%%%%%%%%%%%%%%%%%%%%%%%%%%%
% Figure 6
%%%%%%%%%%%%%%%%%%%%%%%%%%%%%%%%%%%%%%%%%%%%%%%%%%%%%%%%%%%%%%%%%%%%%%
The magenta solid line in Fig.~\ref{fig:Theory} compares the predictions of Eq.~\ref{EqE02} with the experimental results for $\Delta E_\mathrm{IX}$. The model
reproduces reasonably well the measured magnitude and density dependence of the shifts in the regime of low to moderate IX fluid densities (i.e., for IX$_L$ densities below $4-8\times 10^{10}$~$cm^{-2}$). This agreement is quite surprising: Eq.~\ref{EqE02} yields large red-shifts because the expression used for $\omega(k)$ neglects the additional intra-DQW repulsive interactions arising from the polaron density fluctuation, while the IX fluid at this density range ($n_\mathrm{IX}>10^{10}~$cm$^{-2}$) is known to be in a correlated state, where the repulsive interactions play an important role~\cite{Laikhtman_PRB80_195313_09,Shilo_NC4_2335_13,Cohen_NL16_3726_16}. 

\replaced[id=PVSn]
{The increasing role of intra-layer repulsion and  dipolar particle correlations within the IX$_L$ fluid}
{The effect of the increasing role of the intra-layer repulsive interactions and increasing dipolar particle correlations within the IX$_L$ fluid} \cite{Laikhtman_PRB80_195313_09,LozovikSSC07} 
\replaced[id=PVSn]{expresses itself in Fig.~\ref{fig:SpectralResults} as a significant reduction of the energy shifts }{is seen in Fig.~\ref{fig:SpectralResults} as the measured shifts reduce significantly} when the IX$_L$ densities exceed $\sim 8 \times 10^{10}$~cm$^{-2}$. At this density range, the IX$_L$ fluid is \added[id=PVSn]{expected to be} a highly correlated liquid \cite{Laikhtman_PRB80_195313_09,Stern_S343_55_14} 
\replaced[id=PVSn]{ with a linear dispersion relation $\omega(k) \approx c (n_\mathrm{IX})  k$ determined by a speed of sound} 
{and the  dispersion relation should become linear $\omega(k) \approx c (n_\mathrm{IX})  k$, determined by the speed of sound} 
$c (n_\mathrm{IX})$,  which in turn depends on the density $n_\mathrm{IX}$~\cite{LozovikSSC07}.
Under such a linear dispersion, the energy shift becomes:

\begin{equation}
\label{E03}
\Delta E_\mathrm{IX}  =  - n_\mathrm{IX} \mu_1^2 \mu_2^2 \frac{3\pi}{8 L_z^4 m c^2(n_\mathrm{IX})}.
\end{equation}

Numerical computations by Lozovik {\it et al.}~\cite{LozovikSSC07} revealed that \replaced[id=PVSn]{the speed of sound for an IX liquid is given}{for the IX liquid in this regime the speed of sound given} by  $c(n_\mathrm{IX}) \sim c_0 n_\mathrm{IX}^{0.7}$ (cf. Fig.~3b of Ref.~\cite{LozovikSSC07}). In this case, $\Delta E_\mathrm{IX} \sim n_\mathrm{IX}^{-0.4}$ reduces with increasing density. This behavior is reproduced by the thick dotted line in Fig.~\ref{fig:Theory}, which was  determined from Eq.~(\ref{E03}) using the sound velocities from  Ref.~\cite{LozovikSSC07}. It can be shown that the polaron cloud has a gaussian spatial profile \replaced[id=PVSn]{with a gaussian width}
{$~e^{-r^2_\parallel/(2 L^ 2_p)}$ with}   $L_p= 2 L_z/\sqrt{35}$ (cf. Sec.~\ref{DeltaN}). The decreasing energy shifts with increasing $n_\mathrm{IX}$ can \added[id=PVSn]{also} be understood by the increase stiffness of the IX liquid, which results in a smaller polaron density deformation amplitude.

\added[id=PVSn]The polaron binding energies given by Eq.~ \ref{E03} coincides with the reduction of the emission energy of a recombining IX only in the adiabatic approximation, i.e., for interaction processes on a time scale longer than the typical polaron response time,  
$\tau_\mathrm{p}\approx L_p/c(n_\mathrm{IX})=3$~ps for 
$n_\mathrm{IX}=10^{10}$~cm$^{-2}$  and 0.3~ps for $10^{11}$~cm$^{-2}$.
This is a good approximation in view of the long  IX lifetimes.
If, in contrast, the bound single IX recombines within a time shorter then $\tau_p$, it will leave the IX fluid in a deformation state described by a Poissonian superposition of an integer number $n_{ph}=0,1,2,\dots$ of deformation quanta (``phonons''). The characteristic phonon energy can be determined from the Gaussian polaron profile to be   $|\Delta E^\mathrm{na}_\mathrm{IX}|= \sqrt{\pi}\hbar c(n_\mathrm{IX})/(2 L_p)=5.5$~meV for $n_\mathrm{IX}=10^{11}$~cm$^{-2}$ \replaced[id=PVSn]{ thus leading to a red-shift $n_{ph} |\Delta E^\mathrm{na}_\mathrm{IX}|$ for each recombination event (cf. Appendix~\nameref{a:polaron}). 
In this case the red-shift energy of each IX recombination event will be given by $ n_{ph} |\Delta E^\mathrm{na}_\mathrm{IX}|$. }
However, for many such recombination events, and if the linewidth is larger than $|\Delta E^\mathrm{na}_\mathrm{IX}|$, the measured red-shift will be given by their average: $\langle n_{ph}\rangle |\Delta E^\mathrm{na}_\mathrm{IX}|$. Calculating $\langle n_{ph}\rangle$ within the liquid approximation yields an average red-shift energy that differs from that predicted by Eq.~\ref{E03}, up to a numerical factor of order unity (cf. Appendix~\nameref{a:adiabatic}). This shows the robust relation between the polaron binding and the red-shift of the IX emission energies.  

The cross-over from an uncorrelated to a correlated regime should thus significantly reduce the energy shifts at high IX densities. Since the fraction of particles in a correlated state increases with density, one also expects a reduction of $\Delta E_\mathrm{IX}$ at high densities. This behavior agrees with the reduction of the binding energy observed in cw experiments for densities beyond approximately $8\times 10^{10}$~cm$^{-2}$. The polaron model can thus qualitatively reproduce the energy red-shifts over a wide density regime.  \deleted[id=PVSn]{A more sophisticated model is, however, required to account for the details of the density dependence.}

\deleted[id=PVSn]{We note that the above calculated polaron binding energies for a single IX coupled to an IX fluid will coincide with the red-shift of the emission energy of a recombining IX only in the adiabatic approximation, i.e., for interaction processes on a time scale longer than the typical polaron response time  
$\tau_\mathrm{p}\approx L_p/c(n_\mathrm{IX})=3$~ps for 
$n_\mathrm{IX}=10^{10}$~cm$^{-2}$  and 0.3~ps for $10^{11}$~cm$^{-2}$.
This is a good approximation in view of the long  IX lifetimes.
If, in contrast, the bound single IX recombines within a time shorter then $\tau_p$, it will leave the IX fluid in a deformation state described by a Poissonian superposition of an integer number $n_{ph}=0,1,2,\dots$ of deformation quanta (``phonons''). The characteristic phonon energy can be determined from the Gaussian polaron profile to be   $|\Delta E^\mathrm{na}_\mathrm{IX}|= \sqrt{\pi}\hbar c(n_\mathrm{IX})/(2 L_p)=5.5$~meV for $n_\mathrm{IX}=10^{11}$~cm$^{-2}$. In this case the red-shift energy of each IX recombination event will be given by $ n_{ph} |\Delta E^\mathrm{na}_\mathrm{IX}|$. However, for many such recombination events, and if the linewidth is larger than $|\Delta E^\mathrm{na}_\mathrm{IX}|$, the measured red-shift will be given by their average: $\langle n_{ph}\rangle |\Delta E^\mathrm{na}_\mathrm{IX}|$. Calculating $\langle n_{ph}\rangle$ within the liquid approximation yields an average red-shift energy that defers from that predicted by Eq. \ref{E03}, up to a numerical factor of order unity. This shows the robust relation between the polaron binding and the  red-shift of the IX emission energies.}  

\section{CONCLUSIONS}

 We have experimental evidence for the attractive component of the dipolar interaction between IX dipoles in stacked DQWs by spatially-resolved PL spectroscopy. We have shown that the remote interaction between IX fluids located in stacked DQWs leads to changes in the IX spatial distribution as well as to an increase in the IX-IX inter-layer energy $\Delta E_\mathrm{IX}$. Surprisingly, $|\Delta E_\mathrm{IX}|$ values far exceed those expected from the binding of two IXs in a molecule. The magnitude and qualitative density dependence of $|\Delta E_\mathrm{IX}|$ is well accounted for by a many-body dipolar-polaron model. The presented results are expected to challenge state-of-the-art theoretical models of dipolar quantum liquids, however further work will be required to quantify the detailed dependence of the polaron binding energy on IX densities. In particular, it is still not understood why we observe large binding energies, which are qualitatively reproduced by the non-interacting polaron picture of Eq.~\ref{EqE02}, in a density regime where strong intra-layer repulsive interactions are expected to suppress the polaron deformations and hence its binding energies. 
We also note that in the current experiments, the densities of the IX$_H$ fluid were not negligible, therefore the single impurity model used here should be extended in order to get a more quantitative comparison to the experimental data.  
 
The strong attractive inter-DQW coupling opens up possibilities to observe new complex many-body phenomena of dipolar quantum fluids in solid-state systems, that now involve the full anisotropic nature  of the dipole-dipole interactions. Since IX systems can probe density and interaction strengths \replaced[id=PVS]{currently unavailable in }{unavailable in currently available} atomic realizations, it is expected to reveal new collective effects, the attractive dipolar-polaron being a good such example. The sensitivity of the inter-layer coupling to intra-layer fluid correlations demonstrated here can be used as a sensitive tool to \replaced[id=PVS]{probe}{sense} intricate particle correlations in interacting quantum condensates. These experiments also demonstrate the feasibility of dipolar control of inter-layer flow in excitonic devices based on stacked dipolar structures. 
Concepts for the control of IX flows based on repulsive interactions have previously been put forward~\cite{PVS284}. The results presented here enable their extension to attractive potentials, which can be realized using stacked DQW structures.  Finally, the present investigations also open the way for the realization of dipolar lattices in the solid state. One-dimensional lattices can be realized  by simply stacking DQWs. These lattices can be extended to three dimensions by introducing a lateral modulation via electrostatic gates \cite{Remeika_APL100_61103_12,Schinner_PRL110_127403_13} or acoustic fields~\cite{PVS223}.

\section*{ACKNOWLEDGEMENTS}
The authors would like to thank Stefan F\"olsch  and Maxim Khodas for their fruitful discussions and comments on the manuscript.
This research was made possible by the German-Israeli Foundation (GIF) grant agreement: I-1277-303.10/2014 and the Austrian Science Fund (FWF), project number: P29902-N27.

\setcounter{secnumdepth}{0}
\section[A]{APPENDIX A: SAMPLES}
\label{a:Samples}

The studies were carried out in two  (Al,Ga)As layer structures (samples A and B) grown by molecular beam epitaxy on GaAs~(001) at the Paul-Drude-Institut (sample A) and at Princeton University (sample B). Both samples have  DQWs with the same layer structure, as described in the main text.

For sample A, the DQW stack was placed approximately 500~nm away from the semi-transparent top gate and only 100~nm above the bottom electrode. The electric field responsible for IX formation was applied between the top gate and this bottom electrode. The short distance between the DQWs and the bottom electrode minimizes coplanar stray electric fields at the edges of the top gate. This electrode consists of a n-type doped distributed Bragg reflector (DBR) consisting of four Al$_{0.15}$Ga$_{0.85}$As and AlAs layer stacks designed for a central wave length $\lambda_c$~=~820~nm. The DBR enhances the IX emission by back-reflecting the photons emitted towards the substrate. In addition, it suppresses the PL from the substrate (most notably the lines related to the GaAs exciton (around 818~nm) and GaAs:C (830~nm) transitions, which spectrally overlap with the IX PL line. 

In Sample B, the DQW stack was also placed $\approx$520~nm away from the top gate, but was situated 250~nm away from the bottom electrode. The substrate was n-doped and used as the back contact in a Schottky-type diode, with the DQWs being again situated in the intrinsic region. The QWs are GaAs, while the intra-DQW barriers are Al$_{0.3}$Ga$_{0.7}$As. The barriers between substrate and top contact are also Al$_{0.3}$Ga$_{0.7}$As. 

The main difference between the two samples is the addition of a Bragg mirror in sample A, as well as smaller radial electric fields, due to the placement of the DQWs closer to the (semi-infinite) ground plane. The \deleted[id=PVSn]{presence of  the } Bragg mirror allows for the operation of the device at higher electric fields (limited by the breakdown voltage, instead of the photoluminescence flux) so that the IX energies are 30-40~meV lower than the direct excitons. 

\section[B]{Appendix B: DETERMINATION OF IX DENSITY}
\label{a:IXDensity}

\begin{figure}[htbp]
\includegraphics[width =1\columnwidth]{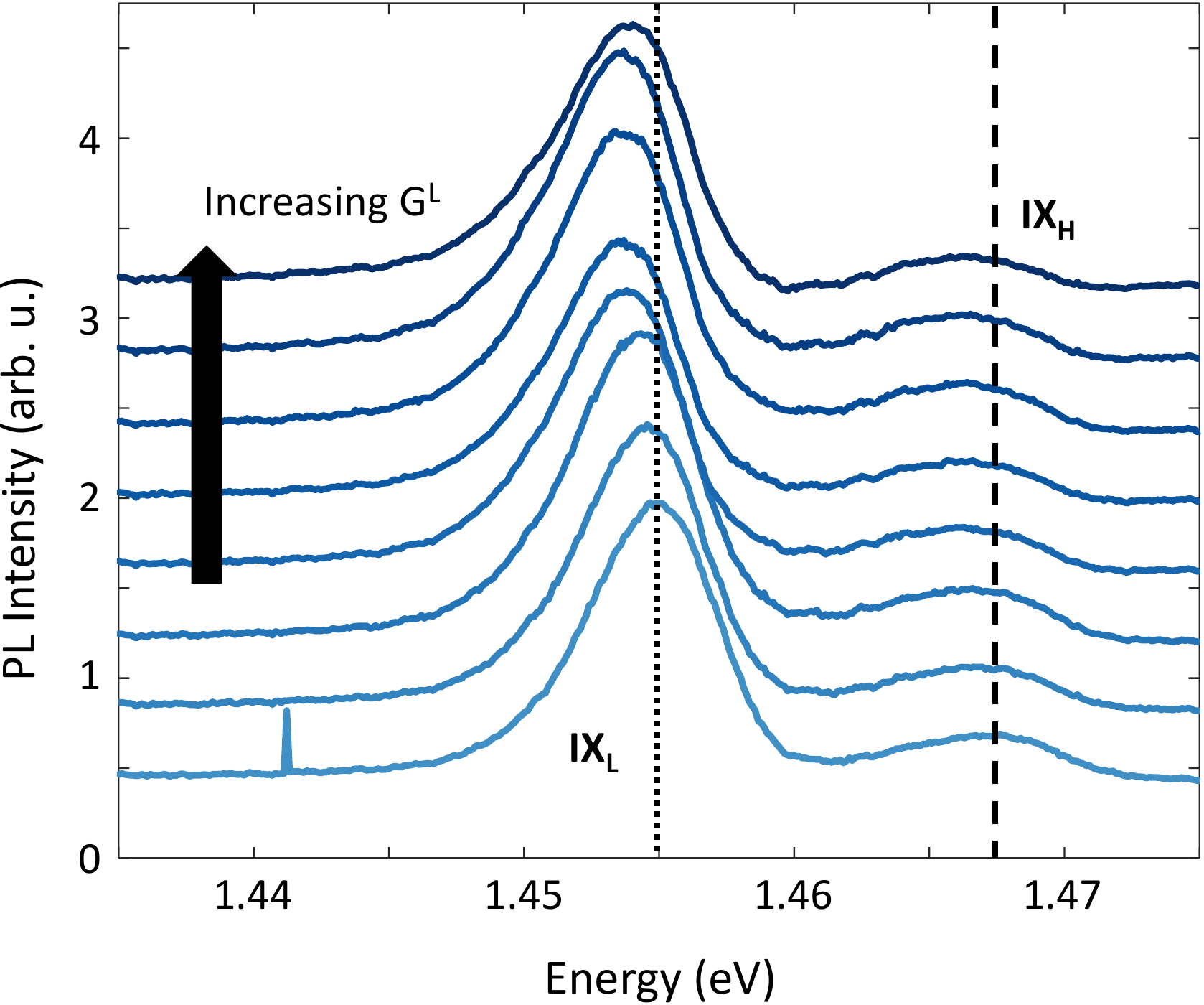}
\caption{ 
Spatially integrated PL spectra for the  IX$_L$ (around 1.455 eV) and IX$_H$ (around 1.468 eV) resonances as a function of the IX$_L$ excitation power G$^L$ (from bottom to top: G$^L$ = 0 (lightest), 10, 20, 30, 43, 55, 80, 130 (darkest)~$\mu$W). The electric field is 38~kV/cm and  G$^H$ = 40~$\mu$W. Data is taken from sample A, but is the equivalent experiment for sample B  shown in Fig.~\ref{fig:SpectralResults} of the main text.  Each integrated spectrum is normalized with respect to its maximum intensity to allow easy comparison of the energetic red shifts with increasing IX$_L$ intensity. The energy of the IX$_L$ and IX$_H$ transitions in the absence of the  G$^L$ excitation are marked by the dot and  dashed lines, respectively. Each of these spectrum would be equivalent to a single data point in Fig.~\ref{fig:SpectralResults}.
}
\label{fig:SMRawData}
\end{figure}

The data shown in Fig.~\ref{fig:SpectralResults} is processed from several individually recorded spectra. The FWHM linewidth of the recorded IX spectra are typically around 2-5~meV, mainly dependent on the density and the integration area. The spectra shown in Fig.~\ref{fig:SMRawData} (recorded from sample A) is typical raw PL data demonstrating the energy shifts induced by the inter-DQW interactions.  The energies used in Fig.~\ref{fig:SpectralResults} (recorded from sample B) were determined from the peak energies and intensities obtained from such spectra using the procedure described below. 

In instances where the diffusion of the IX clouds resulted in pronounced energy shifts, the energy at the highest density was used. 

%%%%%%%%%%%%%%%%%%%%%%%%%%%%%%%%%%%%%%%%%%%%%%%%%%%%%%%%%%%%%%%%%%%%%%%%%%%%%%%%%%%%%%%%%%%%%%
%%%%%%%%%%%%%%%%%%%%%%%%%%%%%%%%%%%%%%%%%%%%%%%%%%%%%%%%%%%%%%%%%%%%%%%%%%%%%%%%%%%%%%%%%%%%%%

The exact calibration of the exciton density is such systems is a well-known challenge. Here we use the following procedure: for every experiment with a given applied bias, we use the experiment with only the $G^L$ laser as a reference. Since this laser creates only a population of IX$_L$, the interactions in this case are only repulsive, leading to a blue shift of the energy with increasing laser power (increasing density). We then choose a point that has an interaction energy well within the range expected for a correlated liquid regime, described in detail in Ref. \cite{Laikhtman_PRB80_195313_09}. We then use Eq. 5.5 in that reference to estimate the density of IX$_L$s for this experimental point. The IX densities of both IX$_L$ and IX$_H$ can then be induced relative to this reference density by comparing the relative emission intensities of each of the IX species to the emission intensity of the reference point, using the procedure developed in Refs. \cite{Shilo_NC4_2335_13,Mazuz-Harpaz_PRB95_155302_17}. Here, it was shown that the emission intensity of the IX, $I_{i}\propto n_i/\tau_i$ where $n_i$ is the IX$_i$ density and $\tau_i$ is the IX lifetime. This lifetime was shown in Ref.~\cite{Mazuz-Harpaz_PRB95_155302_17} to be related to the energy difference between the IX emission and the DX emission energies: 

\begin{equation}
\tau= c_d (\Delta E_\mathrm{DX-IX})^2,
\end{equation}

\noindent where $\Delta E_\mathrm{DX-IX}=E_\mathrm{DX}-E_\mathrm{IX}$ and the  proportionality factor $c_d$ depend on the layer structure of the sample and the applied bias, but does not dependent on the density over a rather wide range of densities. Thus, for every two points with the same applied bias but different laser excitation powers, the ratio between their corresponding IX densities can be found using:

\begin{equation}
\frac{n^{(1)}}{n^{(2)}} =
\frac{\tau^{(1)} I^{(1)}_{PL}} {\tau^{(2)} I^{(2)}_{PL}}
=\frac{I^{(1)}_{PL}}{I^{(2)}_{PL}} \left[\frac{ \Delta E^{(1)}_\mathrm{DX-IX} }{\Delta E^{(2)}_\mathrm{DX-IX}}\right]^2.
\end{equation}

This ratio was used to calibrate the absolute density of all experimental points in any given experiment with a fixed applied bias to the reference point in that experiment.

\section[C]{APPENDIX C: ELECTROSTATIC CONTRIBUTIONS}
\label{a:Electrostatic}

\begin{figure}[htbp]
\includegraphics[width =1\columnwidth]{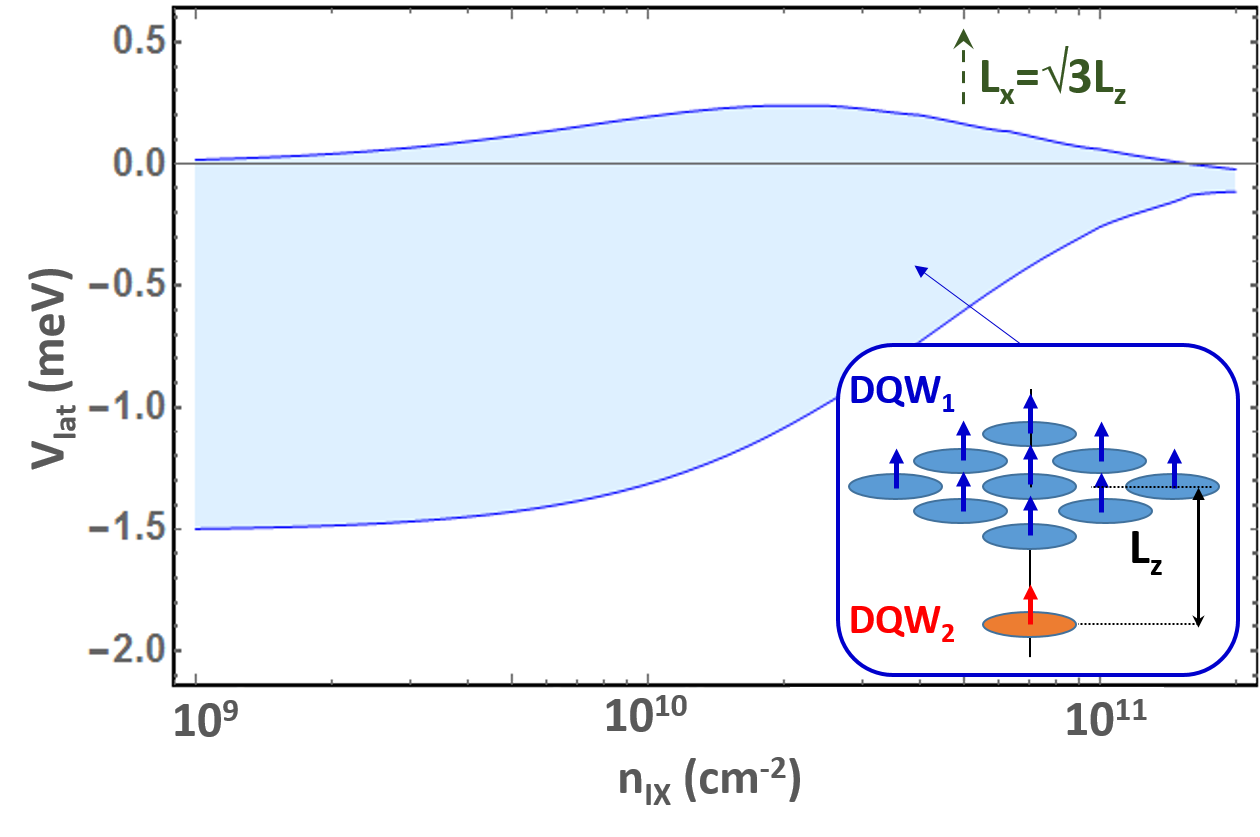}
\caption{ 
Range of energies (shaded region) spanned by the inter-DQW interaction between a single IX in DQW$_2$ and a closed-packed triangular lattice of IXs with density $n_\mathrm{IX}$ in DQW$_1$, as illustrated in the inset.
$n_\mathrm{IX}$ is related to the triangular lattice constant $L_x$ by  $n_\mathrm{IX}=\frac{2}{\sqrt{3} L_x^2 }$.
}
\label{fig:SMTheory}
\end{figure}

The inter-DQW potential in Fig.~\ref{fig:Potential}c of the main text applies for the inter-DQW interaction between two aligned dipoles, each in one of the DQWs. In this section, we have  estimated the dependence of the inter-DQW potential $V_{lat}$ on the density of particles. For that purpose, we calculate the dipolar potential experienced by a single IX in DQW$_2$ due to the coupling  to an excitonic cloud in DQW$_1$ (cf. inset of Fig.~\ref{fig:SMTheory})  by (i) neglecting kinetic effects and (b) assuming that the IXs within the cloud of DQW$_1$ are arranged in a closed-packed triangular lattice with lattice constant $L_x$ \cite{Remeika_CLEO_2_16a}. 

$L_x$, as well as the associated particle density in the triangular lattice $n_\mathrm{IX}=\frac{2}{\sqrt{3} L_x^2}$, are determined by the spot size and intensity of the excitation laser as well as by recombination and expansion rates of the excitonic cloud. 
$V_{lat}(\vec{r})$ was determined by summing the two-particle contributions $U_{dd}(r)$ (cf. Eq.~\ref{Eq1}) over a large number of lattice lattices.  

The shaded region in Fig.~\ref{fig:SMTheory} marks the range of energies spanned by $V_{lat}(\vec{r})$ as the single IX (with coordinate $\vec{r}$) moves relative to the lattice, calculated for different lattice densities. For densities yielding $L_x >>L_z$, the lattice potential around each site resembles the one for $U_{dd}$ in Fig.~\ref{fig:Potential}(c) of the main text and indicated by the dashed horizontal line in Fig.~\ref{fig:Theory}. As  $L_x$ decreases to values comparable to $L_x$, the minima of $V_{lat}(\vec{r})$  remain aligned with the lattice sites. In the opposite limit $L_x << L_z$  $V_{lat}(\vec{r})\rightarrow 0$, thus reproducing the  fact that the electric field generated by an infinite sheet of dipoles vanishes at large distances.   The minimum values for $V_{lat}(\vec{r})$ are always larger than the minimum for the IX-molecule interaction potential $U_{dd}$. This simple  model for the interaction underestimates the  measured binding energies  $|\Delta E_\mathrm{IX}|$ indicated by the symbols in Fig.~\ref{fig:Theory} on the main text.

%%%%%%%%%%%%%%%%%%%%%%%%%%%%%%%%%%%%%%%%%%%%%%%%%%%%%%%%%%%%%%%%%%%%%%%%%%%%%%%%%%%%%%%%%%%%%%
% Figure 6
%%%%%%%%%%%%%%%%%%%%%%%%%%%%%%%%%%%%%%%%%%%%%%%%%%%%%%%%%%%%%%%%%%%%%%%%%%%%%%%%%%%%%%%%%%%%%%

\section[D]{APPENDIX D: POLARON MODEL}
\label{a:polaron}

\paragraph{Two-body interactions.} We consider two layers of excitons with dipole moments $\mu_{1,2} = p_{1,2}/\added[id=ML]{\sqrt{4 \pi \varepsilon \varepsilon_0}} = e d_{1,2}/{\sqrt{4 \pi \varepsilon \varepsilon_0}}$, separated by a distance $L_z$. 
\deleted[id=PVSn]{Let us start by estimating interactions between dipoles in such a configuration.}
 We will treat excitons as point dipoles, which is a good approximation only for $d_{1,2} \ll L_z$. In {our setup} $d_{1,2}/L_z \sim 0.3$, however, it should {still provide}  a reasonable estimate. The dipole-dipole interaction between the dipole $\mu_{1}$ (located at $\rho=0$ in DQW$_1$) and the dipole $\mu_{2}$ (located in DQW$_2$ at a lateral separation $\rho$)  \added[id=PVSn]{can be written as (cf.~Eq.~1)}:
\begin{equation}
\label{Vr}
%V(r) = \frac{\mu_1 \mu_2}{(L_z^2 + r^2)^{3/2}}\left(1 - \frac{3L_z^2}{L_z^2 + r^2} \right)
V(\rho) = \frac{\mu_1 \mu_2}{(L_z^2 + \rho^2)^{3/2}}\left(1 - \frac{3L_z^2}{L_z^2 + \rho^2} \right)
\end{equation}
This interaction is sign-changing, so a net mean-field interaction of a dipole with a dipolar plane vanishes:
\begin{equation}
\label{Vrint}
\int d^2 \rho V(\rho) = 0.
\end{equation}

In order to solve Eq.~\ref{H1} of the main text we first note that the Fourier transform of the two-body interaction potential of  Eq.~\eqref{Vr} can be expressed as:

\begin{equation}
\label{EqVk}
% V(k) = \int d^2 r V(r) e^{- i \mathbf{k r}} = - \mu_1 \mu_2~2 \pi k~e^{-k L_z}
 V(k) = \int d^2 \rho V(\rho) e^{- i \mathbf{k \rho}} = - \mu_1 \mu_2~2 \pi k~e^{-k L_z}
\end{equation}  

\noindent In addition,  $f(k) = \left[n_\mathrm{IX} \varepsilon(k)/(\hbar \omega(k))\right]^{1/2}$ is a function that depends on the density $n_\mathrm{IX}$, single-particle energy $\varepsilon(k) = \hbar^2 k^2/(2m)$ as well as on the  correlation state of the IX gas expressed in terms of its dispersion relation $\hbar\omega(k)$. 
%$m = m_e + m_{hh}$ is the exciton mass. 

If we consider a static impurity (an ``infinite-mass polaron'', $M=\infty$, located at $\vec{r} = 0$), the Hamiltonian~\eqref{H1} can be diagonalized using a coherent-state transformation

\begin{equation}
\hat S = \exp \left[- \sum_\vec{k} \frac{U(k)}{\added[id=PVS]{\hbar}  \omega(k)} \left(\bed_\vec{k} - \be_\vec{k}  \right) \right],
\end{equation} 

\noindent which gives the following ground-state energy shift:

\begin{equation}
\label{E0}
\Delta E  =  - \sum_\vec{k} \frac{U(k)^2}{\added[id=PVS]{\hbar}\omega(k)} 
\end{equation} 
(the ground state is given by $\ket{\psi}= \hat S \ket{0}$).

 One can see that $\Delta E $ is always negative: this is a general property of Hamiltonians with linear coupling, such as Eq.~\eqref{H1}.  The energy shifts for a gas of non-interacting excitons expressed by Eq.~{EqE02} of the main text was obtained by integrating Eq.~\ref{E0} using  the dispersion relation is given by $\hbar \omega(k) \equiv \varepsilon(k) $. The corresponding expression for an interacting exciton gas (Eq.~\ref{E03} of the main text) was determined in the same way using a dispersion relation $\omega(k) \approx c (n_\mathrm{IX})   k$, where $c (n_\mathrm{IX})$ is the density dependent speed of sound.
 
\section[E]{APPENDIX E: POLARON DENSITY PROFILES}
 
In real space, the density deformation of DQW$_1$ is given by $\Delta n_\mathrm{IX} (\rho) = \bra{\psi} \bed_\vec{r}\be_\vec{r} \ket{\psi}$, where $\bed_\vec{r} = \int d^2 k/(2\pi)^2 \bed_\vec{k} e^{i \vec{k} \vec{\rho}}$\deleted[id=PVSn]{, where $\vec{\rho}=\vec{r_\parallel}$}. 
\note{[ML: here $r$ is actually $r_\parallel$, but the latter doesn't look nice as a variable.]} In the case of a correlated excitons in DQW$_1$, the density deformation can be approximated by a Gaussian at small values of $\rho$:

\begin{equation}
\label{DeltaN}
\Delta n_\mathrm{IX} (\rho)  =  n_\mathrm{IX} \frac{ \mu_1^2 \mu_2^2}{2 {\added[id=PVS]\hbar} m c^3(n_\mathrm{IX})}   \frac{9\pi}{16 L_z^5 } e^{-\frac{\rho^2}{2 L_p^2}},
\end{equation}

\noindent where $L_p = 2 L_z/\sqrt{35}$.  
By integrating $\Delta n_\mathrm{IX}$ over the DQW plane one obtains a total density excess corresponding to approx. 0.1 particles for $ n_\mathrm{IX}=10^{10}$~cm$^{-2}$.

\section[F]{APPENDIX F: NON-ADIABATIC ENERGY SHIFTS}
\label{a:adiabatic}

The polaron wavefunction $\ket{\psi}=\sum_q F({\bf q})\bed_\vec{r} \ket{0}=\sum_q F({\bf q}) \ket{\bf q}$, where $F(\bf q)$ is the Fourier transform of the gaussian real space profile with width $L_p$ (cf. Eq.~\ref{DeltaN}). For a state having a single polaron quantum (``phonon'' ), the normalization condition $\sum_{\bf q' ,q''}\bra{\bf q'} F^*(\bf q')F(\bf q'')\ket{\bf q''}$ yields $F({\bf q})=\sqrt{8 m L^2_p}e^{-\frac{\rho^2}{2 L_p^2}}$. The single phonon energy can be determined by replacing $\omega({\bf q}) \approx c (n_\mathrm{IX}) q$ in the following expression:

\begin{equation}
|\Delta E^\mathrm{na}_\mathrm{IX}|=\bra{\psi} \hbar\omega({\bf q})   \ket{\psi}= \frac{\sqrt{\pi}\hbar c(n_\mathrm{IX})}{2 L_p}.
\end{equation}

\noindent This expression yields $|\Delta E^\mathrm{na}_\mathrm{IX}|= 1.1 $~meV for $n_\mathrm{IX}=10^{10}$~cm$^{-2}$ and 5.5~ meV for  $n_\mathrm{IX}=10^{11}$~cm$^{-2}$.

%%%
%The previous expressions for the energy shifts in the correlated regime were obtained in the adiabatic approximation, i.e., for interaction processes on a time scale longer than the typical polaron response time  $\tau_\mathrm{p}\approx L_p/c(n_\mathrm{IX})=3$~ps for $n_\mathrm{IX}=10^{10}$~cm$^{-2}$ and  0.3~ps for $n_\mathrm{IX}=10^{11}$~cm$^{-2}$. This is a good approximation in view of the long  IX lifetimes. If, in contrast, the bound single IX recombines within a time shorter then $\tau_p$, it will leave the IX fluid in a coherent deformation state   described by a Poissonian superposition of an integer number $n_{ph}=0,1,2,\dots$ of deformation quanta (``phonons''). The characteristic phonon energy can be estimated from sound velocity and the width $2L_p$ of the Gaussian polaron profile to be   $|\Delta E^\mathrm{na}_\mathrm{IX}|= \sqrt{\pi}\hbar c(n_\mathrm{IX})/(2 L_p)= 1.1 (5.5)$~meV for $n_\mathrm{IX}=10^{10}$~cm$^{-2}$ ($10^{11}$~cm$^{-2}$).
%

The average phonon energy, which corresponds to the red-shift and broadening of a bound IX in the  non-adiabatic approximation,  can then be calculated according to:%The energy of the emitted IX will be red-shifted and broadened  by approximately 

\begin{align}
\langle n_{ph} \rangle |\Delta E^\mathrm{na}_\mathrm{IX}|&= 
|\Delta E^\mathrm{na}_\mathrm{IX}| \int {d\rho^2  \Delta n_\mathrm{IX} (\rho)} \\
&=  - f^\mathrm{na}   n_\mathrm{IX} \mu_1^2 \mu_2^2 \frac{3 \pi}{8 L_z^4 m c^2(n_\mathrm{IX})} .
\end{align}

\noindent This expression is similar to Eq.~\ref{E03} of the main text, exception for a pre-factor  $ f^\mathrm{na} ={ 3\pi^{3/2} }{/4} \sim 1.4$.  The density-dependent shifts are comparable to the ones determined in the adiabatic approximation and, thus, much smaller than the measured ones.

\bibliographystyle{apsrev4-1}
\bibliography{mypapers,literature}
%\bibliography{word}

\clearpage

\end{document}